\documentclass[twocolumn,superscriptaddress]{revtex4}
\usepackage{graphicx}
\usepackage{epstopdf}
\usepackage{amsmath}

\begin{document}
\draft
\title{Slow dynamics in a primitive tetrahedral network model}

 \author{Cristiano De Michele} \affiliation{ {Dipartimento di Fisica and
  INFM-CRS-SOFT, Universit\`a di Roma {\em La Sapienza}, Piazzale A. Moro
  2, 00185 Roma, Italy} }
  \author{Piero Tartaglia} \affiliation{
  {Dipartimento di Fisica and INFM-CRS-SMC, Universit\`a di Roma {\em
  La Sapienza}, Piazzale A. Moro 2, 00185 Roma, Italy} }
\author{Francesco Sciortino} \affiliation{ {Dipartimento di Fisica and
  INFM-CRS-SOFT, Universit\`a di Roma {\em La Sapienza}, Piazzale A. Moro
  2, 00185 Roma, Italy} }         
\begin{abstract}
We report extensive Monte Carlo and event-driven molecular dynamics simulations of the fluid and liquid phase of a primitive model for silica recently introduced by  Ford, Auerbach and Monson [J. Chem. Phys. {\bf 17}, 8415 (2004)].
We evaluate the iso-diffusivity lines in the temperature-density plane to
provide an indication of the shape of the glass transition line.  Except for large
densities, arrest is driven by the onset of the tetrahedral bonding pattern and the
resulting dynamics is {\it strong} in the Angell's classification scheme. We compare structural  and dynamic properties with  corresponding results of two recently studied primitive models of  network forming liquids ---
a primitive  model  for water  and a angular-constraint free model of  four-coordinated particles ---  to  pin down the role of the geometric constraints associated to the bonding. Eventually we discuss the similarities between "glass" formation in network forming liquids and "gel" formation in  colloidal dispersions 
of patchy particles.
\end{abstract}
\maketitle
\section{Introduction}

Primitive models for atomic and molecular liquids have been and currently are
very valuable in the study of the structure and of the thermodynamic properties of 
several compounds\cite{rpm,Nez01a,Nez05a,2000JPCM12R411T}.  In most cases, primitive models condense the inter-particle repulsion in a hard-core potential, while attractive interactions are modeled as simple square well potentials. An ingenious choice of the hard-core diameters and of the location and number of the square-well sites generates models which are able to reproduce many of the essential features of the liquid (and sometime of the crystal\cite{Veg98a,For04a}) states, despite their intrinsic simplicity. From a theoretical point of view, primitive models are particularly useful, since they allow for a close comparison between theoretical predictions and numerical "exact" results.  It is not a coincidence that several novel approaches to the physics of liquids (including critical phenomena and liquid-solid first order phase transitions) are first tested against these models before being extended to more complicated continuous potentials.

 Primitive models are relevant not only in the study of liquids, but  more generally in the study of colloidal systems, especially when the solvent properties are neglected and colloidal particles are represented as units interacting via an effective potential\cite{Lik01b}.  Also in colloidal physics,
despite the severe approximations, simple hard-sphere and square well models\cite{Fof02a,Zac02a,STadvmodphys} have been of significant utility in deepening our understanding of the essence of the equilibrium liquid state and of the glass transition. The relevance of these models to colloidal physics is expected to grow in the
near future, when the newly synthesized  
colloidal particles with patterns of sticky patches on their surfaces
will be produced in large quantities\cite{Manoh_03,Zerro_05,Mohw2005}. 

A recent line of investigation\cite{Dem06a}  has called attention on the similarities between glass formation
in network forming liquids  and gels made by particles with limited valency\cite{Zac05a,Mor05c,Zac06a}.  It has been suggested that the progressive shrinking of the unstable liquid-gas region which takes place on
progressively decreasing the maximum number of bonds\cite{Zac05a,Bia06}  is a necessary condition for observing
arrest at low packing fraction  in the absence of phase separation.   Particles interacting only with attractive spherical potentials (beside the hard-core repulsion) can not form "equilibrium" ideal gels\cite{Fof05aPRL,idealgel}, since phase separation prevents the possibility of reaching in an homogeneous state low $T$, where the lifetime of the bond  would be sufficiently long to provide a finite elasticity to the structure.  Interestingly enough, network forming liquids fall in the category of limited-valency. Accordingly, in network forming liquids the slowing down on approaching the glass transition  can be interpreted as  the molecular counterpart of the gelation process in colloidal systems with limited valence.  It has also been suggested that the bond-energy sets a well defined energy scale in all microscopic processes and induces an Arrhenius dependence of the $T$-dependence of all characteristic times\cite{Mor05c}.  Limited valence gels, similarly to network forming liquids, are expected to be strong in Angell's classification\cite{Ang91a}.

To deepen our understanding of this intriguing hypothesis, it is useful to go back to primitive
models of network forming liquids and explore their dynamic properties in relation to  the fields of thermodynamic stability.  Indeed, understanding the loci of dynamic arrest,  i.e. the regions in phase space where disordered arrested states can be expected, and the competition between arrest in a non ergodic disordered structure as compared to crystallization in an ordered  one, is central  in  any modeling of material properties.  It is a fortunate that recently, Ford, Auerbach and Monson\cite{For04a} have introduced a simple model  based on low coordination and strong association for silica, one of the most important network forming material. This primitive model for silica (PMS)  envisions a silicon atom as a hard-sphere, whose surface is decorated by four sites, arranged according to a tetrahedral geometry. The oxygen atom is also model as a hard-sphere, but with only two additional sites. The only (square well) attraction takes place between distinct sites of Si and O atoms. Despite the crude modeling, the resulting phase diagram --- which includes three solid phases, corresponding to cristobalite, quartz and coesite, a gas and a fluid phase --- compares very favorably with the experimental one\cite{For04a}. 

In this article we study static and dynamic properties of this model  in the fluid and liquid phase,  by performing an extensive set of Monte Carlo (MC) and event-driven molecular dynamics (ED-MD) simulations. Beside structural and thermodynamic properties, we estimate the line of dynamic arrest in the temperature $T$- density $n$ plane. 
With extensive simulations, we equilibrate the system down to temperatures where a full tetrahedral structure develops and most of the particles are fully bonded.  
When possible, we compare the novel  PMS  results with corresponding data for two primitive models of four-coordinated particles: a primitive model for water (PMW), introduced by Nezbeda and coworkers\cite{Kol87a}  and recently revisited \cite{Dem06a};  a limited-valence square-well model $N_{max}$, introduced
by Speedy and Debenedetti\cite{Spe94aMP,Spe95aMP,Spe96aMP} and recently studied as a model for network forming liquids or as a model for gels in  a series of papers\cite{Zac05a,Mor05c,Zac06a,Mor06a,Zac06cJPCM}. 

The trend coming out from such a comparison 
sheds light on the role of the geometric constraints (introduced by bonding)  and on how the coupling between bonding and local density may provide a  driving force favoring a liquid-liquid phase separation\cite{Poo92a,Poo05JPCM}.

\section{The Model and Numerical Details}
The PMS is a rigid site model in which Si and O atoms (in a 1 to 2 ratio) are modeled as hard spheres complemented by   additional sites located at fixed distances  from the particle centers. 
The hard-core of the Si-Si repulsive interaction  is $\sigma_{SiSi}\equiv \sigma$, where $\sigma$
defines the length scale. The O-O hard core is defined by $\sigma_{OO}=1.6 \sigma$, while
the Si-O repulsive interaction is again of hard sphere type but with range $\sigma_{SiO}=\sigma$.  
In this respect the model defines a non-additive hard-sphere mixture. 
The Si particle is decorated by 
four additional sites located along the direction of a tetrahedral geometry on the surface of the hard sphere, i.e. at distance $0.5 \sigma$ from the
particle center. Each O atom  is  associated to two additional sites $S_1$ and $S_2$ located at distance $0.5 \sigma$ from the particle center $C$.  The angle $S_1 \hat C S_2$ is fixed to $145.8 \deg$.

The only attractive interaction takes place between pairs of  Si and O sites. 
Sites interact  via a square well (SW) potential $u_{SW}$, i.e.
\begin{eqnarray}
u_{SW}=-u_0~~r<\delta \\ 
\nonumber
~~~~~=0~~~~~~r>\delta,
\end{eqnarray}
defined by a small well width  $\delta= (1-\sqrt{3}/2) \sigma \approx 0.134 \sigma$. The depth of the square well potential $u_0$ defines the energy scale.    Pairs of  Si-O particles with pair interaction energy equal to $-u_0$ are  considered bonded.
The value of $\delta$ selected in Ref.\cite{For04a} is  slightly larger than the  $0.5 ( \sqrt{5-2 \sqrt{3}}-1)\sigma \approx 0.119\sigma$ value which would guarantees 
the absence of double bonding at the same site due to steric constraints.  
If the value of $\delta$ is chosen in such a way that each site can be
engaged at most in one bond, the (negative of the) potential energy of the system  $E$ coincides, configuration by configuration,  with the number  of bonds $N_b$.   Similarly, the fully bonded state  coincides with the ground state of the system in which each
Si in involved in exactly four bonds.  Hence the ground state energy is
$E_{gs}=-4$ per Si particle. In the present model, the choice of $\delta$ would in principle allow for double bonding at the same site, as alluded  previously. Despite this theoretical  possibility, we have checked that this rare event of
double bonding never takes place in the present study and hence the system
potential  energy and  (the negative of ) the total number of bonds are always coincident.

We have studied a system of $N=216$ particles with periodic boundary conditions for several  values of the volume $V$, spanning a wide range of 
number densities $n \equiv N/V$ and temperatures  $T$, where $T$ is measured in units of $u_0$ ($k_B=1$).   For $n \le 0.15$, we have studied a system of $N=2160$ particles, to better follow the expected onset of critical fluctuations.
We perform both MC and ED-MD. In MC, we define a  step as  an attempt to move 
each of the $N$ particles.  A move is defined as a  translation  in each direction of a random quantity distributed uniformly between $\pm 0.05~\sigma$ and a rotation around a random axis of random angle distributed uniformly 
between $\pm 0.5$ radiant.   Equilibration was performed with MC, and monitored via the evolution of the potential energy (a direct measure of the number of bonds in the system).  The mean square displacement (MSD) was calculated to guarantee that each particle has diffused in average more than its diameter. In evaluating the MSD we have taken care of subtracting the center of mass displacement, an important correction in the low $T$ long MC calculations. Indeed, at low $T$ simulations required more than $5\times 10^9$ MC steps, corresponding to several months of CPU time for each investigated state point.  

We have also performed ED-MD simulations of the same system,  starting from configurations equilibrated with MC. The algorithm implemented to calculate the trajectory of a system of particles interacting by  an hard-core complemented by one or more site-site square well potentials is described in details in Ref.\cite{Dem06a}. 
The only difference with respect to the algorithm illustrated in Ref.\cite{Dem06a} 
is that here 3 different linked lists have been used for the 3 possible interactions: Si-Si, Si-O, O-O.
This is necessary in order to guarantee a good computational efficiency since 
$\sigma_{OO}=1.6 \sigma > \sigma$.
In ED-MD, particle masses have been set to $m_{Si}=14$ and $m_{O}=8$. The moments of inertia have been assumed to have the following values $I_{Si}=14$ and $I_{O}=8$. Time is measured in units of $(m_O \sigma^2 / 8u_0)^{1/2} $.

We conclude this section by briefly describing the PMW\cite{Kol87a} and $N_{max}$\cite{Zac05a} models. Both models are one-component models. The PMW has a strong tetrahedral character, imposed by the tetrahedral arrangement of four SW sites. Two of the sites mimic the H atom and the two other sites the two lone-pairs. The    $N_{max}$ model is an isotropic square well model with an additional constraint  on the maximum number of bonds. A model with  four-coordinated particles is generated by imposing that the maximum number of
particles  probing the SW potential is limited to four.  A fully bonded particles has
four neighbors arranged in a tetrahedral geometry in the PMW and 
in a  random geometry in the $N_{max}$.

\section{Results: Statics}
\subsection{Potential Energy}

\begin{figure}[tbh]
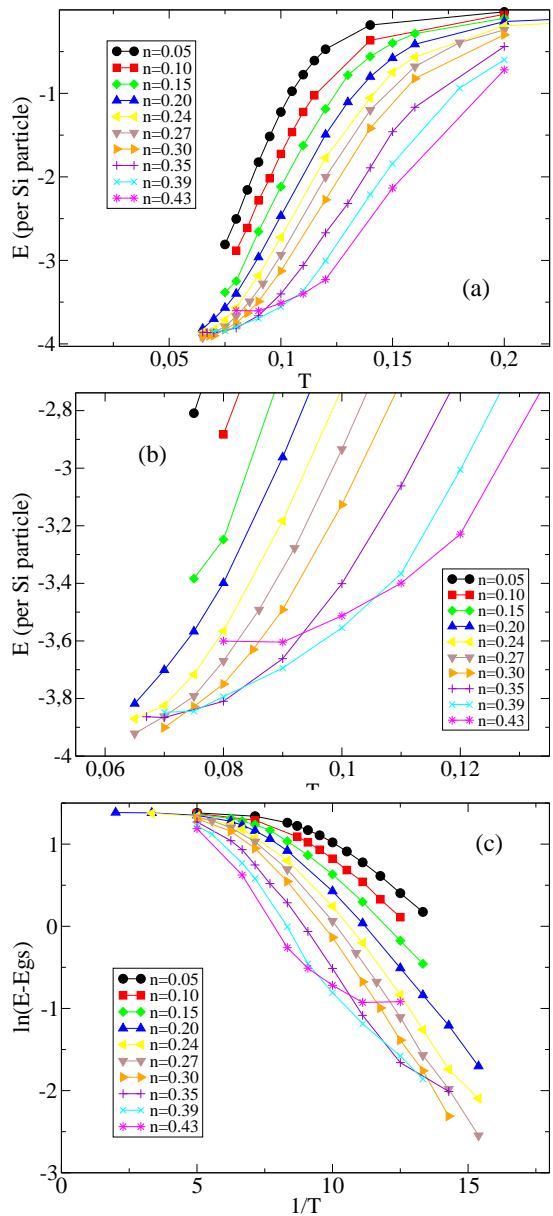

\centering
\includegraphics[width=0.4\textwidth]{EvsTMC}
\includegraphics[width=0.4\textwidth]{EinsetvsTMC}
\includegraphics[width=0.4\textwidth]{EvsusuTMC}
\caption{(a) Potential energy per Si particle $E$ vs.  temperature $T$ for different isochores. Panel (b) shows an enlargement of the low $T$ part to highlight the approach to the ground state energy $E_{gs}=-4$.  Panel (c) shows $E-E_{gs}$ as a function of $1/T$ to show that in the optimal network density region (see text) the behavior of  $E-E_{gs}$ shows an Arrhenius law. }
\label{fig:evsT}
\end{figure}

The average potential energy of the system per Si particle $E$, is shown in Fig.~\ref{fig:evsT}. For $n \lesssim 0.15$,  lines stop at $T=0.075$, the lowest $T$ at which the fluid is stable. Indeed, for lower $T$, liquid-gas separation takes place, as detected by a progressive growth of the structure factor peak at low $q$. For $n \gtrsim 0.20$,  phase separation is never observed  in the  range of temperatures where it is possible to achieve equilibration with present computational resources.
The enlargement of the low $T$ region  (Fig.~\ref{fig:evsT}b) shows that 
in the region of densities $0.24 \lesssim n \lesssim 0.30$ the potential
energy approaches the ground state energy
$E_{gs}=-4$,  suggesting that a fully connected network of bonds develops
on cooling.   Outside the range  $0.24 \lesssim n \lesssim 0.30$,  as $T \rightarrow 0$,  the energy does not appear to approach $E_{gs}$.  
The region $0.24 \lesssim n \lesssim 0.30$ defines the optimal network density region\cite{Dem06a},  i.e. the range of $n$ values where a fully  bonded state can be reached at low $T$.    For densities lower or higher than the optimal one, the formation of a fully connected network is hampered by geometric constraints: at lower $n$, the large inter-particle distance   acts against the possibility of forming a fully connected network, while at large $n$, packing constraints promoting close packing configurations are inconsistent with the tetrahedral bonding geometry.   These features introduce a significant coupling between the maximum number of bonds (and hence energy)  and density.  It is interesting to observe that, 
compared to the previously studied PMW case\cite{Dem06a}, the region of optimal densities is wider in PMS, due to the fact that the bond is mediated by the O particles and that the bonding sites on the O particles are not located along a
diameter.



\begin{figure}[tbh]
\centering
\includegraphics[width=0.5\textwidth]{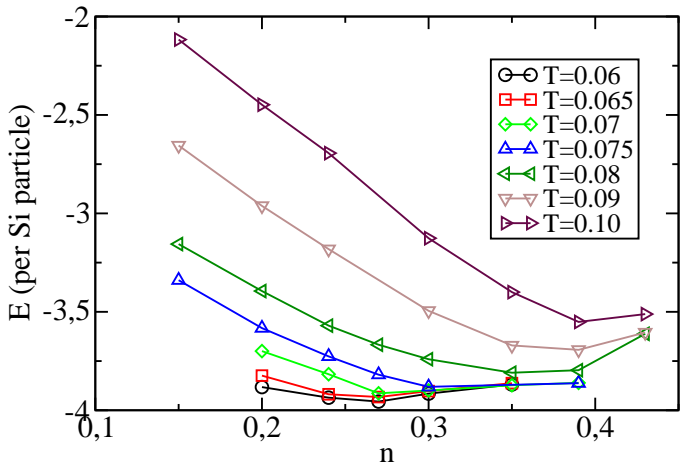}
\caption{Potential energy  per Si particle vs. $n$ along isotherms. Note the minimum that develops  around $n \approx 0.27$ at low $T$.}
\label{fig:enevsrho}
\end{figure}

The possibility of reaching values the ground state energy in the optimal network density region suggests that indeed, the "equilibrium" state at $T=0$ is a disordered fully connected network. In the above sentence, the word "equilibrium" is meant to stress that the system ---while being in metastable equilibrium with respect to the crystal state --- manages to equilibrate in the restricted part of configuration space associated to liquid microstates. Support for this hypothesis comes from the  $T$ dependence of the energy  at low $T$.  As shown in Fig.~\ref{fig:enevsrho}c, in the region $0.24 \lesssim n \lesssim 0.30$ the energy per Si particle is very well described by the Arrhenius law $E-E_{gs} \sim e^{-E_a/T}$, where $E_a \approx 0.5$ is the activation energy and $E_{gs}=-4$ is the $T=0$ limit.   The validity of the Arrhenius law for the energy is characteristic of strong liquids and  it has been observed also in simulations of more sophisticated models of silica\cite{Hor99a} and water\cite{Poo05JPCM}. Differently from the case of continuous 
potentials, primitive models do not require any fitting of the ground state energy. It is important to stress that,  despite the  disorder intrinsic in the liquid state, it appears still possible to fully satisfy the bonds in a disordered homogeneous structure, i.e. it is possible to reach the perfect random tetrahedral continuous network\cite{Zac32a} continuously cooling the liquid.

The validity of the Arrhenius law has been verified also in the case of other models\cite{Zac05a,Zac06a,Dem06a}. It is worth recalling that an Arrhenius law for the potential energy with $E_a=0.5$ is also a prediction of the Wertheim's thermodynamic perturbation theory\cite{Wer84a,Wer84b}, a theory developed under the assumption of completely uncorrelated bonding sites.   While in the
 PMS  and in the $N_{max}$ model\cite{Zac05a} 
$E_a \approx 0.5$, in the  PMW case $E_a \approx 1$. The introduction of the O particle as a bond-mediator appears to provide sufficient flexibility in bonding to release the strong geometric constraint imposed by the strict tetrahedral geometry of the PMW.

Fig.~\ref{fig:enevsrho} shows the $n$ dependence of $E$ along isotherms, to highlight the presence of the minimum in the density dependence of 
the energy which develops at low $T$.  The minimum is located in the
region   $0.24 \lesssim n \lesssim 0.30$ where density allows for the development of a fully bonded tetrahedral network.   Fig.~\ref{fig:enevsrho} also shows
that a region of negative curvature in $E(n)|_T$ develops just outside the
optimal density region, suggesting that the formation of the network
and the coupling between density and energy may act as a driving force
for a thermodynamic destabilization of  the liquid, favoring the possibility of
a liquid-liquid phase transition\cite{Poo92a,Poo05JPCM}.    

It is interesting to compare the density in this optimal region with the
density of a diamond crystal. This comparison can be done defining a
reduced scaled density $n_s$, as the ratio between the number of
four-coordinated particles (the sites of the network) and the volume measured in units  of  bond-distance $d_b$.  In these units, the scaled density for the  close packed diamond structure is $n_s^{diamond} \approx 0.65$.
From the parameters of the PMS, the expected bond distance between two Si particles bonded via a common O particle is approximatively $d_b=1.91\sigma$.   If one focuses only on the Si atoms (considering the oxygen only as bond-mediator)  then the network sites number density is reduced by a factor of 3 as compared to the system $n$.  Then the range  $0.24 \lesssim n \lesssim 0.30$ converts to a scaled density  $n_s \equiv  \frac{n}{3} \cdot (\frac{d_b}{\sigma})^3 \approx 2.32 n$ in the region $  0.56 \lesssim n_s \lesssim 0.69$,
suggesting that the region of optimal network densities is located around the diamond crystal density. 

\subsection{Structure}

\begin{figure}[tbh]
\centering
\includegraphics[width=0.49\textwidth]{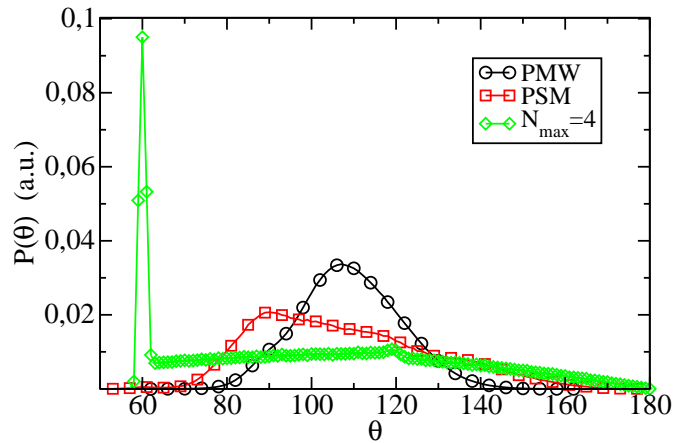}
\caption{Distribution of the angle $\theta$ between connected triplets of particles for the  PMS ,  PMW  and $N_{max}=4$ models. In the case of  PMS , $\theta$ is the Si-Si-Si angle, in the case of 
PMW and $N_{max}=4$  models $\theta$ is the particle-particle-particle angle.}
\label{fig:angle}
\end{figure}

We start by focusing on the geometry of the Si tetrahedral network. Fig.~\ref{fig:angle} shows the Si-Si-Si angle ($\theta$) distribution at a state point in the optimal density region at low $T$.
It has been recognized in the past that this distribution provides
relevant information on the topological properties of
tetrahedral networks and helps quantifying the quality of the tetrahedral structure\cite{1997PhRvL..78.1484M,2000PhRvB..62.4985B}. For the  PMS , the distribution extends from  70 to 160 degrees, peaking around 90 degrees. The
average value is $\bar \theta =107.7$ and a variance $\sigma_\theta=20.5$. 
It is instructive to compare the  PMS  distribution, with the distribution found in the
PMW and $N_{max}$ models, two related primitive models for four coordinated networks. 
In the case of PMW, the distribution is highly peaked, with $\bar \theta =109$ and
 $\sigma_\theta= 12.0$, in agreement with the highly directional character of the
 interactions. The opposite $N_{max}$ case is described by a very wide angle distribution,
 with  $\bar \theta =100$ and  $\sigma_\theta= 32.$. Differently from the  PMS  and  PMW  models,  in the  $N_{max}$ model configurations with touching triplets of spheres ($\theta \approx 60 $)  are also found,  since angular constraints are missing.

\begin{figure}[tbh]
\centering
\includegraphics[width=0.4\textwidth]{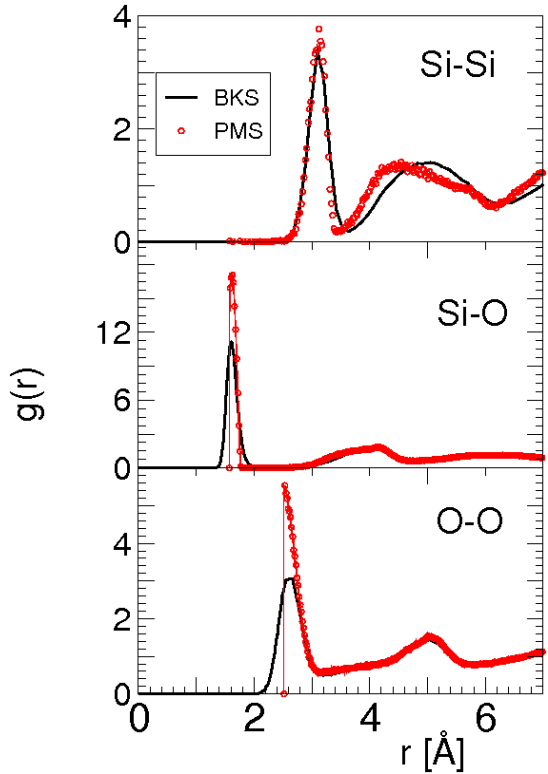}
\caption{Radial distribution functions  $g(r)$ for Si-Si, O-O and Si-O at $n=0.30$ and $T=0.07$ (symbols). Lines are results for BKS silica\cite{bks} from Ref.~\protect\cite{Hor99a} at $T=2750 K$. Note that, in performing the comparison,  $\sigma$ has been fixed to $\sigma = 1.574 \AA$ 
}
\label{fig:grr030}
\end{figure}

\begin{figure}[tbh]
\centering
\includegraphics[width=0.45\textwidth]{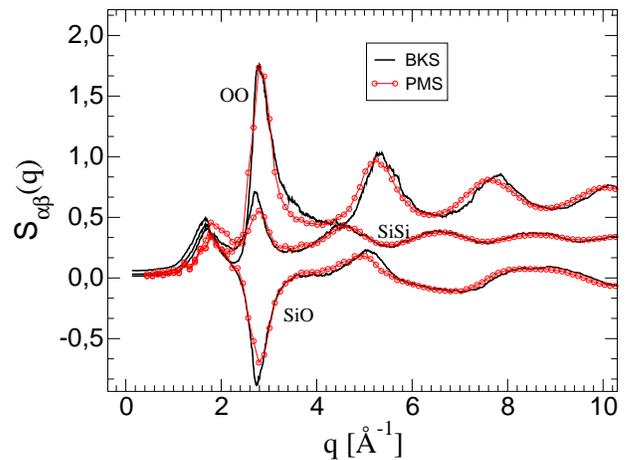}
\caption{Partial structure factors  $S_{\alpha,\beta}$ (with $\alpha$ and $\beta$ labeling the particle type [Si,O])   at $n=0.30$ and $T=0.07$. Lines are results for BKS silica from Ref.~\protect\cite{Hor99a} at $T=2750 K$.
Note that, in performing the comparison,  $\sigma$ has been fixed to $\sigma = 1.572 \AA$ }
\label{fig:sqr030}
\end{figure}

While our aim is not the general comparison between the present model and silica, for which more
realistic potentials are available\cite{bks}, it is interesting  to  look at the structure of the system in the optimal network region and compare it with the structure generated by more realistic potentials.  Detailed information on the structure of the system on cooling are contained in the particle-particle  radial distribution function $g_{\alpha,\beta}(r)$ and in its Fourier transform, the particle-particle structure factor $S_{\alpha,\beta}(q)$ (where $\alpha$ and $\beta$ label the different particle species). Fig.~\ref{fig:grr030} shows the three partial radial distribution functions and compares them with data from Ref.~\protect\cite{Hor99a}. The corresponding comparison in $q$-space is shown in Fig.~\ref{fig:sqr030}. It appears that the  PMS  is able to capture the structure of the tetrahedral network  to a good extent. This also suggests that simulations with the  PMS  can be used as a tool for generating highly bonded configurations  which could be used as input for simulation based on realistic potentials.

\section{Dynamics}

In this section we focus on dynamic properties of the model, and in particular on the
dynamics of Si particles, to put an emphasis on the dynamics of the tetrahedral network.
In this respect, the O particles provide the bonds between the nodes of the Si network.

\subsection{Mean Square Displacement}

\begin{figure}[tbh]
\centering
\includegraphics[width=0.45\textwidth]{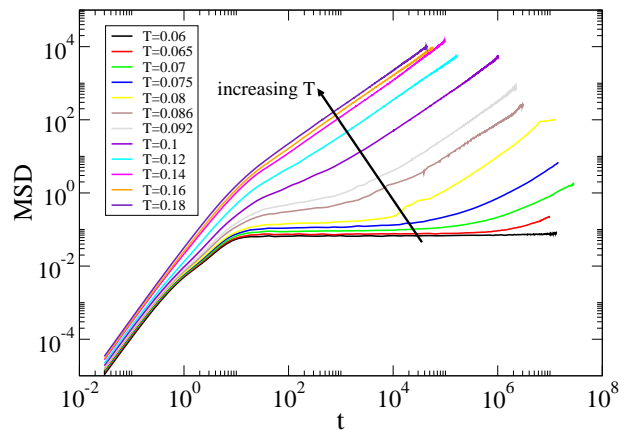}
\caption{MSD vs. $T$  for Si particles  at $n=0.27$.}
\label{fig:msd}
\end{figure}

Figure \ref{fig:msd} shows the mean square displacement (MSD) of the Si particles at one selected $n$.  The behavior of the MSD is characteristic of systems approaching an arrested states. The MSD shows an initial ballistic region, followed a low $T$ by a plateau and, for very long times, a crossover to a diffusive dependence.   It is interesting to observe that the height of the plateau, which can be operatively defined as the point in which the MSD vs $\ln(t)$ curve changes concavity, decreases on $T$. This suggests that,
in the studied interval, the localization length associated to the single particle dynamics   changes with $T$.  In the present model, the restriction of the cage cannot be associated to a thermally induced decrease of the vibrational amplitudes (as is the case of continuous potentials) and thus signals an effective reduction of the explored volume in the cage due to a  progressive development of a bonding pattern.   

\subsection{Diffusion Coefficient}

\begin{figure}[tbh]
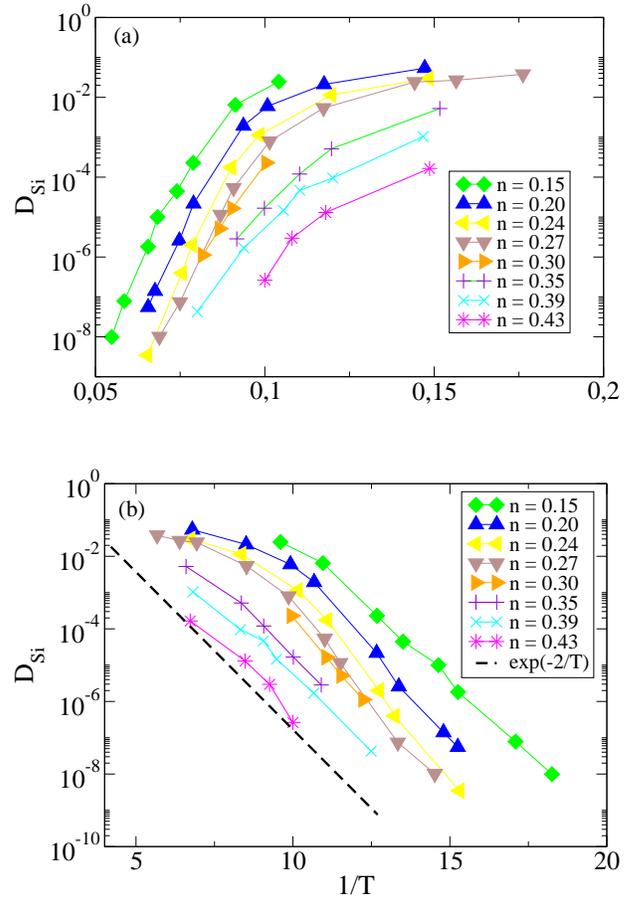

\centering
\includegraphics[width=0.45\textwidth]{diffA}
\vskip 0.50cm
\includegraphics[width=0.45\textwidth]{diffAvsusuT}
\caption{Temperature dependence of the diffusion coefficient of the Si particles  $D_{Si}$  at all studied densities.  (a)   $D_{Si}$ vs T. (b) $D_{Si}$ vs 1/T.}
\label{fig:diffAB}
\end{figure}

The MSD long time limit provides a measure of the diffusion coefficient $D$.  Fig.~\ref{fig:diffAB} shows that with present numerical resources, the slowing down of the dynamics in this model can be studied over more than six orders of magnitude.  While at high $T$  diffusion becomes $T$ independent (an evidence that $n$ is the only relevant variable at high $T$), on cooling $D(T)$ crosses  to a $T$ dependence consistent with an activated law. The linearity of $\log(D)$ vs $1/T$ is very striking in the region of the optimal number densities, where an unconstrained tetrahedral network can form at low $T$.  The activation energy of the diffusion process is close to $2.5$.  In agreement with
previously investigated models of network liquids\cite{Hor99a,   Voi01a,Poo05JPCM}, slow dynamics in the optimal network
region appears to be controlled by activated processes. This suggests that strong-liquid
behavior, in Angell's classification scheme, is indeed intimately connected to the
limited valency of the inter particle interaction potential.    
Comparing with the PMW data, no diffusional anomalies are observed, i.e. diffusion
along isotherms is always monotonically decreasing on increasing $n$. In the  PMW  case,  at low $T$  both diffusion minima and diffusion maxima have been observed. This difference can be understood in terms of structural properties and related to the fact that the distribution of tetrahedral angles in PMS is much wider than the corresponding distribution in PMW.

\subsection{Isodiffusivity Lines} 

\begin{figure}[tbh]
\centering
\includegraphics[width=0.45\textwidth]{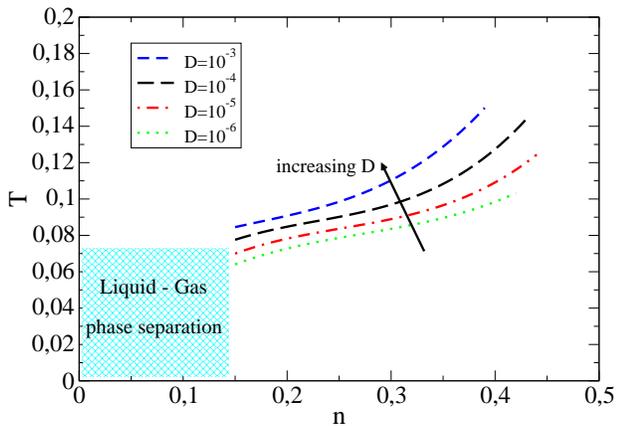}
\caption{Isodiffusivity line for Si particles, for four different values of $D_{Si}$, separated by one decade each. The shaded area schematically indicates the region where liquid-gas phase separation is observed.}
\label{fig:isodiff}
\end{figure}

A possible operational way to determine the shape of the dynamic arrest lines
is provided by a plot of the isodiffusivity curves in the $(T,n)$ phase diagram\cite{Fof02a,Zac02a}.
For the case of Si particles, these lines are shown in 
Fig.~\ref{fig:isodiff}, for three
different $D$ values, separated by one order of magnitude each. The isodiffusivity curves start on the left from the gas-liquid spinodal (i.e. can not be extended to lower
densities due to the presence of the unstable region in the phase diagram)  and  after running essentially parallel to the $T$ axis, start to cross toward a packing controlled
dependence. Again, the simplicity of the model does not leave any ambiguity
that,  at sufficiently large $T$,  the HS dynamics will be recovered. 
 Hence, isodiffusivity curves will become essentially vertical at larger $T$.

The shape of the isodiffusivity lines in the present model  is similar to the one observed in
other models of tetrahedrally coordinated particles. The lowest density at which 
arrest is possible (in an homogeneous, non phase-separated system) is fixed by the boundary of the spinodal curve. In this respect, this appears to be a common feature of all liquids, independently from the maximum valence. On the other end, it is important to realize that the valence (the number of possible nearest neighbors attractive interactions)  controls the location of the high-density side of the spinodal curve.  Only  when the valence of the particles is reduced below six and the spinodal has moved to small packing fractions,  the bond-driven  arrest  line becomes accessible and arrest via activated dynamics can be observed\cite{Zac05a}. 

\section{Discussion and Conclusions}

The aim of the work is to frame the dynamics observed in the primitive model for silica with respect to recent studies of the essential ingredients for slowing down and arrest at low packing fractions and of the connection between glass transition in network forming liquids and gel formation in colloidal particles with patchy interactions.  Indeed, patchy colloidal particles of new generation\cite{Manoh_03,Zerro_05,Mohw2005} may be designed in such a way to closely resemble the primitive model investigated in this and in previous works.
The PMS  interpolates between two  recently investigated models for tetrahedral network forming liquids, the PMW\cite{Dem06a}  and the $N_{max}$\cite{Zac05a} models, which were selected as the extreme cases of highly spatially-correlated  directional bonding  and completely uncorrelated bonding directions. In the case of  PMS , the action of the O particle as mediator of the Si-Si bonding makes it possible to retain  the tetrahedral geometry of the Si-Si bonding with an intermediate constraint on the values of the Si-Si-Si angles.

\begin{figure}[tbh]
\centering
\includegraphics[width=0.45\textwidth]{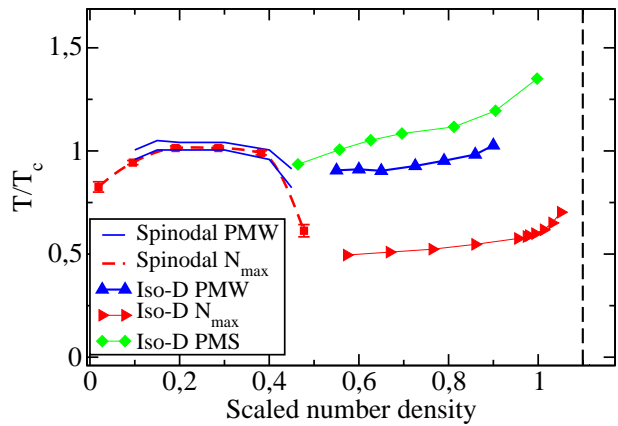}
\caption{Comparison between three primitive models for tetrahedral network forming liquids. The dashed line delimitates the HS glass transition at $\phi=0.58$. The three $T_c$ values are 0.22 ($N_{max}$), 0.1095 (PMW), 0.075 ( PMS ). The scaled density $n_s$ (defined as number of four-coordinated particles divided the volume measured in units of the particle-particle distance) coincides with the density in the case of PMW and $N_{max}$  and it is  2.32 times the  PMS  number density. }
\label{fig:phase}
\end{figure}

In all three models of four-coordinate particles, liquid-gas phase separation is observed only at low $n$, below  the density at which the 
open tetrahedral structure of the fully connected network is observed.  The three
models can be compared in the $T$-density plane using the scaled density 
previously defined.   A comparison of the phase diagram of the three potentials in this scaled unit is shown in Fig.~\ref{fig:phase}.  For all potentials,
the liquid-gas phase separation region is confined to the case $n_s \lesssim 0.5$.  The figure also shows three iso-diffusivity lines, the smallest available iso-diffusivity line for each of the three models. The shape of the isodiffusivity lines is also very similar, confirming that in all these models the bond energy scale controls the slowing down  between the spinodal and the  excluded-volume region.

\begin{figure}[tbh]
\centering
\includegraphics[width=0.45\textwidth]{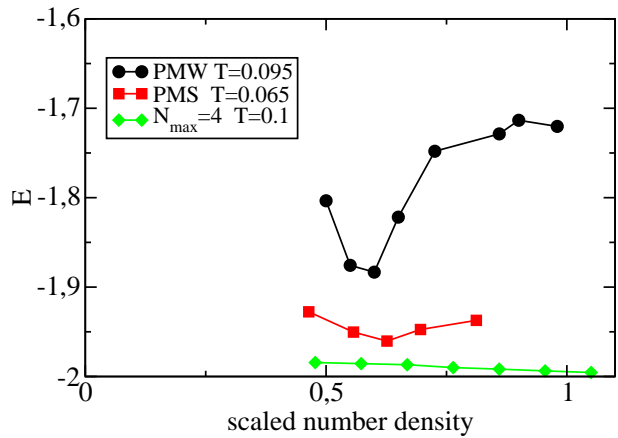}
\caption{Density dependence of the potential energy  per network particle for the three primitive models. The energy for PMS has been scaled by a factor of $T$ since the Si-Si bond requires two bonds via the O particle. The scaled density coincides with the density in the case of PMW and $N_{max}$ and it is  2.32 times the number density for the  PMS .}
\label{fig:evari}
\end{figure}

The coupling between density and extensive bonding  has an interesting  consequence with respect to the disputed  possibility of a liquid-liquid instability, in addition to the gas-liquid phase separation. It has been suggested that a liquid-liquid separation, between two disordered liquid states differing in density
(and hence in possible bonding patterns)  could be a generic feature of all network forming materials with a significant correlation induced by the geometry of the bonds\cite{PooleSci97}. Also in this respect it is interesting to compare the behavior of the three models, i.e.  the behavior of the potential energy per network particle as a function of  $n_s$  at low $T$, shown in  Fig.\ref{fig:evari}. For each potential, the lowest $T$ at which equilibration was achieved is reported.  It is striking to see how pronounced it the minimum in $E(n)|_T$  for the PMW case as compared to the weak minimum 
displayed by the PMS and by the absence of any minimum in $N_{max}$.
The strength of the minimum correlates with  the variance of the  distribution of the tetrahedral network angles. If the onset of a fully bonded network is possible only
for a small range of angles, then the approach  of the lowest possible energy
state is possible only in a very restricted range of densities. For the case of PMW, where a strong geometric constraint among the bonding sites is present, the region of densities where  an almost fully bonded state is reached is localized in a small window around $n_s=0.6$.   In the two other cases, where bonds are rather flexible in orientation, fully bonded states are  technically possible in a much wider window of densities.  Results shown in Fig.~\ref{fig:evari}  thus suggest that the possibility of generating a convex $n$ dependence of the energy
 (and hence a destabilizing contribution to the free energy of the system and the possibility of a liquid-liquid critical point)  is intimately connected to the bonding geometry, since only when the angular bonding energy is sufficiently restricted then  full bonding requires a well defined optimal  density, generating a strong coupling between energy and  density.


An interesting question is posed by the leading arrest mechanism  in different parts of the
phase diagram. While there is no doubt about the fact that at large $n$ packing constraints
(or more generally the shape of the repulsive potential) controls arrest and the fact that instead
the bonding energy controls arrest at low packing fraction, it is still unclear if
the two arrest mechanisms crossover continuously from one to the other (in which case
we would be entitled to call glasses all arrested states) or if the two arrest mechanisms
are intrinsically different and require different names (in which case one would be tempted to 
define a glass an arrested state at large $n$ and a $gel$ the arrested state at small $n$).
From the simulation, it appears rather clearly that 
(i) the slowing down of the dynamics at small $n$ is characterized by activated dynamics, i.e. it is characteristic of  strong glass-forming liquids.
(ii) that slowing down of the dynamics along isochores can be divided into three parts: 
a constant high $T$ part, a crossover intermediate part and an Arrhenius low T part. 
The existence of a low $T$ Arrhenius part has the profound consequence that, technically,
the arrest line is located at $T=0$. If this is the case, then one would rather imaging that
packing arrest line meets the $T=0$ bond arrest line at a finite $n$, providing support to the
possibility of a discontinuous cross-over.  
On the other end, it is important to recall the evidence presented in the literature of the possibility of interpreting the cross-over dynamics observed in more realistic models of tetrahedral forming liquids as the BKS model for silica\cite{Hor99a,Voi01a} and the SPC/E model for water\cite{Sci96b,Sci97b,Sta99a,Fab99b} in term of ideal mode-coupling theory.  While a MCT comparison is impracticable for the present model, due to the non-sphericity of the interaction potential,  the cited results for BKS and SPC/E support the possibility that, at least the cross-over from packing to bonding can be captured by a unique theoretical approach.

A final consideration concerns the possibility of calculating in a numerical exact way, along the lines of Ref.~\cite{Mor05c,Mor06a}, the configurational entropy of this model, especially around the optimal density, where the fully bonded state is almost reached within the simulation time. A comparison between the estimate based on the BKS  potential\cite{Voi01a,2006cond.mat..1259S} and the  PMS 
can help resolving the functional form of the density of states and the connections between low-temperature activated dynamics and landscape properties\cite{2005JSMTE..05..015S}. Work in this direction is underway.

\section{Acknowledgements}
We thank E. Zaccarelli and N. Mosseau for discussions. We acknowledge support from MIUR-COFIN, MIUR-FIRB and CRTN-CT-2003-504712.

\bibliographystyle{./apsrev}
\bibliography{./fileunico,./altri,./new}
                                 
\end{document}